\documentclass[aps,prl,preprint,groupedaddress,showpacs]{revtex4}
\usepackage{epsfig}
\begin{document}

\title{Terahertz frequency standard based on three-photon coherent population trapping}


\author{C. Champenois}
\email[]{caroline.champenois@univ-provence.fr}
\author{G. Hagel}
\author{M. Houssin}
\author{M. Knoop}
\author{C. Zumsteg}
\author{F. Vedel}
\affiliation{Physique
des Interactions Ioniques et Mol\'eculaires (CNRS UMR 6633),
Universit\'e de Provence, Centre de Saint J\'er\^ome, Case C21,
13397 Marseille Cedex 20, France}

\date{\today}

\begin{abstract}
A scheme for a THz frequency standard based on three-photon
 coherent population trapping in stored ions is proposed. Assuming the propagation directions of the three lasers obey the phase matching condition, we show that stability of few  10$^{-14}$ at one second can be reached with a precision  limited by power broadening to $10^{-11}$ in the less favorable case. The referenced  THz signal  can be propagated  over long distances, the useful information being carried by the relative frequency of the three optical photons.
\end{abstract}


\maketitle

Thanks to the progress made in the generation of continuous THz radiation, Terahertz spectroscopy is strongly gaining in interest in different
domains, in particular for imaging in astronomy and biology and for
material analysis  \cite{kawase04}. In this paper we propose a frequency
standard in the Terahertz domain, based on a three-photon dark
resonance  where the three wavelengths belong to the visible domain. Atomic frequency standards in the optical domain can now reach
performances in the 10$^{-16}$ range, thanks to laser-cooling of the
atoms or single ions and to progress made in the stabilization of the clock lasers 
 \cite{oskay06,boyd07}. The proposed THz frequency standard aims at a 
lower relative precision of about 10$^{-11}$, which can be realized in a robust
 and easily achievable experimental configuration and with a large atomic sample, allowing
 frequency stabilities  of a few parts in 10$^{14}$ at one second.

In the scheme described here,  the THz reference signal is produced by coherent population
trapping (CPT) in a three-photon process \cite{champenois06} involving  visible lasers in a
Doppler-free configuration \cite{grynberg76}. For instance, this process can occur in several ions
commonly used for optical metrology or quantum computation. For these atoms (Ca$^+$,
Sr$^+$, Ba$^+$, Hg$^+$), the narrow dark line resulting from CPT corresponds to the magnetic dipole transition between the $ n D_{3/2}$
and $ n D_{5/2}$ metastable states, whose frequency lies  in the THz range.  This
electric-dipole forbidden line has been studied and probed directly in former experiments \cite{werth89,madej92,siemsen92}, but in a configuration sensitive to the Doppler effect. The proposed experiment can be seen as  analogous to two-photon CPT  microwave cesium clock  \cite{zanon05} where  a trade-off is made between robustness and clock precision. Two-photon CPT is also considered for an  optical clock based on neutral $^{88}$Sr \cite{santra05} and a three-photon coherent
process has been proposed as an optical frequency standard in
alkaline earth atoms in \cite{hong05}.

In this paper, we first introduce the three-photon coherent
process resulting in the sharp line that can serve as a frequency
standard. This process is largely discussed in  \cite{champenois06} and we mention
 only the points useful for metrological application. Then, the
expected performances of such a standard are studied to show a possible stability  better than $8 \times 10^{-14}$ at one second.

\begin{figure}[htb] \begin{center}
\epsfig{file=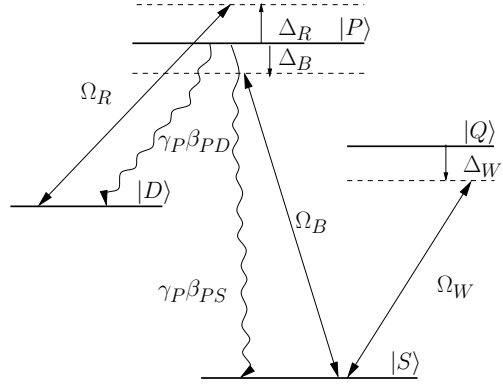, height=5.cm}
\caption{$N$ Level scheme: The states $|D\rangle$,
$|P\rangle$, $|S\rangle$ coupled by laser couplings $\Omega_R$ and $\Omega_B$ 
form a $\Lambda$- configuration, state
$|S\rangle$ couples weakly to the metastable state $|Q\rangle$ by $\Omega_W$.
The wavy lines indicate the
radiative decay. Parameters and possible atomic species are
discussed in the text. \label{fig_N}} \end{center}
 \end{figure}

 The scheme we propose can apply to any atomic system  composed of four electronic levels
which are coupled by laser fields, according to the $N$-shaped
scheme depicted in Fig.~\ref{fig_N} and where states $|S\rangle$,
$|D\rangle$ and $|Q\rangle$ are (meta)stable while state $|P\rangle$ is
short-lived and decays radiatively into $|S\rangle$ and $|D\rangle$.
This  level configuration  is realized, for instance, in
alkaline-earth atoms with hyperfine structure and in alkali-like
ions with a metastable $d$-orbital, such as  Hg$^+$,  Ba$^+$,
Sr$^+$, or Ca$^+$. In this manuscript we focus on this last case  where the
 levels can be identified with the states
$|S\rangle=|S_{1/2}\rangle$, $|P\rangle=|P_{1/2}\rangle$,
$|D\rangle=|D_{3/2}\rangle$, and $|Q\rangle=|D_{5/2}\rangle$. The dark line can be observed by collecting the photon signal on the $|S\rangle \to |P\rangle$ transition. Here, all the transitions are electric-dipole allowed except  $|S_{1/2}\rangle \leftrightarrow |D_{5/2}\rangle$ (named  $|S\rangle \leftrightarrow|Q\rangle$  in
Fig.~\ref{fig_N}), which is an electric quadrupole transition with a
linewidth of the order of one Hertz. The magnetic-dipole transition  $|D_{3/2}\rangle \leftrightarrow |D_{5/2}\rangle$ ($|D\rangle \leftrightarrow |Q\rangle$) considered as the frequency standard has a spontaneous emission rate  of the order of $10^{-6}$ s$^{-1}$ which can be neglected.

For our purpose, the evolution of the internal degrees of freedom of a motionless atom can be understood in the dressed state picture, where the non-coupled eigenstates are defined by
 \begin{equation}
 H_0 =  \hbar\Delta_R|D\rangle\langle D|+\hbar\Delta_B|S\rangle\langle S|+\hbar(\Delta_B-\Delta_W)|Q\rangle\langle Q|
 \end{equation}
with detunings defined as $\Delta_B=\omega_B-\omega_{PS}$, $\Delta_R=\omega_R-\omega_{PD}$, and
$\Delta_W=\omega_W-\omega_{QS}$ ($B, R, W$ stands respectively for the $|S\rangle \leftrightarrow |P\rangle$, $|D\rangle \leftrightarrow |P\rangle$ and $|S\rangle \leftrightarrow |Q\rangle$ transitions. $\omega_X$ is the laser frequency on the transition labelled $X$ and $\omega_{IJ}$ is the frequency of the atomic transition $|I\rangle \leftrightarrow |J\rangle$). $\Omega_{R,B,W}$ are the corresponding Rabi frequencies characterizing the laser couplings. The radiative processes taken into account
couple $|P\rangle$ to states $|S\rangle$ and $|D\rangle$ by the decay rate $\gamma_P$  and
branching ratio $\beta_{PS}/\beta_{PD}$ ($\beta_{PS}+\beta_{PD}=1$). The radiative decay of states $|Q\rangle$ and  $|D\rangle$, whose lifetime is of the order
of 1 s is neglected in the analytical model we
present below. It should  be noted that  the dipole and
the quadrupole couplings differ by a few orders of magnitude. Nevertheless, as the process described in this manuscript is of interest when
 state $|Q\rangle$ is weakly coupled to the
$\Lambda$-scheme ($|S\rangle$, $|P\rangle$, $|D\rangle$, coupled by the two strong lasers) it is thus experimentally feasible with
standard laser sources.

\begin{figure}[htb] \begin{center}
\epsfig{file=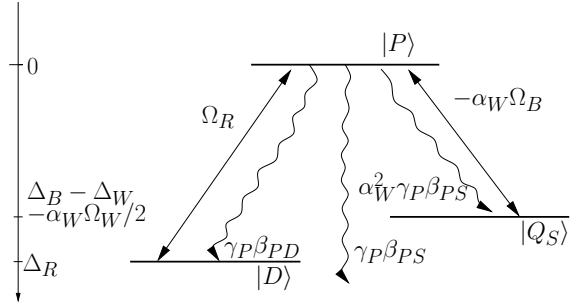, height=4.cm}
\caption{Part of the dressed state picture relevant for the three-photon resonance condition, giving rise to the coherent  dark state. The straight lines stands for laser couplings and the wavy lines for radiative decay (see text for details). \label{fig_lambda}} \end{center}
 \end{figure}

 Considering that $|Q\rangle$ is weakly coupled to   $|S\rangle$, the subsystem ($|S\rangle$,  $|Q\rangle$) can be diagonalised and solved analytically to first order in $\alpha_W=\Omega_W/2\Delta_W \ll 1$. The new eigenstates are then 
\begin{eqnarray}
\left|S_Q\right> & =&{\mathcal N} \left(|S\rangle+\alpha_W|Q\rangle\right)\\
\left|Q_S\right> & =&{\mathcal N} 
\left(|Q\rangle-\alpha_W|S\rangle\right) \nonumber \end{eqnarray} 
(where ${\mathcal N}$ is the normalization factor) with
eigenfrequencies light-shifted by $\pm \alpha_W\Omega_W/2$. $\left|Q_S\right>$ being coupled to
 $\left|P\right>$ by the Rabi frequency $-\alpha_W\Omega_B$, the dressed states configuration ends up in a $\Lambda$-scheme (formed by $\left|Q_S\right>$,$\left|P\right>$ and $\left|D\right>$, see Fig.\ref{fig_lambda}), well known to give rise to coherent population trapping into a dark state when  the metastable states  $\left|Q_S\right>$ and $\left|D\right>$ are degenerated when dressed by the laser photons \cite{arimondo96}. This condition is fulfilled on  the light-shifted three-photon resonance condition
\begin{equation}
\label{3-photon}\Delta_{\mathit{eff}}=\Delta_R+\Delta_W-\Delta_B+\alpha_W \Omega_W/2=0. 
\end{equation}
  Detuned from the two-photon resonance condition ({\it i. e.} $\Delta_R-\Delta_B \ne 0$, see \cite{champenois06} for justification), the atomic system is then pumped into the dark state
 \begin{equation}
|\Psi_{Dark}\rangle ={\cal N}'\left({\cal E}
|D\rangle+|Q_S\rangle\right) \label{psiD} 
\end{equation}
with ${\cal E}=\alpha_W\Omega_B/\Omega_R$ and normalization
factor ${\cal N}'$. 

 The sharp line produced in the fluorescence spectra by pumping into the dark state  can be used to assure $\Delta_R+\Delta_W-\Delta_B+\alpha_W\Omega_W/2=0 $. Experimentally, this is realised by locking one of the three lasers on the dark line. For the following laser frequencies, it translates into
\begin{equation}
\omega_R+\omega_W-\omega_B=\omega_{QD}-\alpha_W\Omega_W/2.
\label{accord}
\end{equation}
Provided the light-shift $\alpha_W\Omega_W/2$ is made negligible or controlled, the frequency combination $\omega_R+\omega_W-\omega_B$ can be referenced to the magnetic-dipole transition frequency $\omega_{QD}$ which lies between 1.8 THz for Ca$^{+}$ and 451 THz for Hg$^{+}$ (see table \ref{tab_lambda}).
 \begin{table}
 \caption{Wavelength involved in the three-photon process and frequency of the reference transition\label{tab_lambda}}
 \begin{ruledtabular}
 \begin{tabular}{ccccc}
Ions  & B & R & W & $D_{3/2}-D_{5/2}$   \\
\hline
Ca$^+$ & 397 nm & 866 nm & 729 nm & 1.82 THz \\
Sr$^+$ & 422 nm & 1092 nm & 674 nm & 8.40 THz \\
Ba$^+$ & 493 nm & 650 nm & 1761 nm & 24.0 THz  \\
Hg$^+$ & 194 nm & 10.67 $\mu$m & 282 nm & 451 THz\\
 \end{tabular}
 \end{ruledtabular}
 \end{table}

The stability of the dark state requires the reduction of the  relative phase diffusion of the three lasers to a diffusion rate smaller than the dark line width. This can be achieved by reducing the spectral width of each laser or, as a less stringent condition, by phase-locking the three lasers on the same reference like a frequency comb \cite{udem99}.  The fundamental limit for the linewidth of the coherent dark line, defined by the lifetime of the metastable states,  may not be reached, mainly because of power broadening, which is known to be one of the main limiting factor in two-photon CPT clock \cite{knappe05}. For evaluating the dark line width,  the system can be approximated by the $\Lambda$-scheme depicted in Fig.\ref{fig_lambda} where the decay from $|P\rangle$ to $|Q_S\rangle$  is $\alpha_W^2 \gamma_P \beta_{PS}$. Because of spontaneous emission from  $|P\rangle$ to  $|S_Q\rangle$, the  decoherence rate between  $|P\rangle$ and $|Q_S\rangle$ or $|D\rangle$  is equal to  $\gamma_P/2$. An analytical expression of the linewidth of coherent dark line can be found in \cite{these_zanon} for very general cases. Such an expression is simplified when, for $\alpha_W^2 \ll 1$ and $\alpha_W^2\Omega_B^2 \ll \Omega_R^2$ one has $\beta_{PD} \Omega_B^2 \gg \beta_{PS} \Omega_R^2$, which is obeyed for realistic working parameters. Then, if the decoherence sources like the laser relative phase diffusion are neglected, the dark linewidth (FWHM) can be approximated by
\begin{equation}\label{power}
\Gamma_{\mathit{eff}}=\frac{\Omega_R^2}{\gamma_P\sqrt{1+\left(3-\frac{\gamma_{\Lambda}}{\gamma_P}\right)\frac{\Omega_R^2}{\gamma_P^2 \beta_{PD}}+\frac{4\Delta_R^2}{\gamma_P^2}   }}
\end{equation}
with $\gamma_\Lambda=\gamma_P(\beta_{PD}+\alpha_W^2\beta_{PS})$. When collected on the $|P\rangle \to |S\rangle$ transition, the fluorescence rate emitted around the dark resonance $\Delta_{\mathit{eff}}=0$ (Eq. \ref{3-photon}) scales with
\begin{equation}\label{signal}
{\cal S}=\gamma_P\frac{\Gamma_{\mathit{eff}}^2}{\Omega_R^2}\frac{\beta_{PS}}{\beta_{PD}}.
\end{equation}
According to these last two equations, the linewidth can be tuned continuously by $\Omega_R$ and in practice, results from a tradeoff between the desired signal over noise ratio $S/N= \sqrt{\eta{\cal S}/2}$ (with $\eta$ the detection efficiency, $1/2$ indicating we work at half maximum and if we assume no background)  and signal linewidth $\Delta \ge \Gamma_{\mathit{eff}}$. Both parameters control the stability of the frequency standard, as can be clearly seen in the Allan deviation
\begin{equation}
\sigma_y(\tau)=\frac{\Delta}{\omega_{QD} (S/N)}\sqrt{\frac{T_c}{\tau}}
\end{equation}
which  quantifies the stability versus the integration time $\tau$, $T_c$ being the cycle time.  Increasing the $S/N$ ratio can be realised without power broadening the transition by increasing the number of ions emitting the fluorescence signal. For ions in radiofrequency trap, this option is rarely used because of the  Doppler effect due to the micromotion (the radiofrequency driven motion), which increases with the number of ions \cite{dehmelt69,wineland88}. In the case of a three-photon resonance, there is a (first order)Doppler-free configuration \cite{grynberg76} which is defined by the phase matching condition
\begin{equation}
\Delta {\bf k}={\bf k_R}-{\bf k_B}+{\bf k_W}={\bf 0}.
\end{equation}
The driven micromotion can not be laser cooled and its contribution to the second-order Doppler effect can  not be nulled.  Assuming the thermal macromotion (motion in the pseudo-potential created by the RF-field) is laser cooled, the relative shift induced by second order Doppler effect can be estimated \cite{wineland88} from the number $N_i$ of trapped ions, their mass $M$ in atomic units and the secular frequency of oscillation $\nu_S$ (MHz):
\begin{equation}
\Delta f_{D2}/f_0=-3.0 \times 10^{-14}(\nu_S N_i/M)^{2/3}.
\end{equation}
In a centimeter-size trap, where $N_i=10^5$ and $\nu_S=0.1$ MHz are typical values, this relative shift is $-1.2\times 10^{-12}$ for calcium ions, the lightest of the species considered here. For this ion, the resulting frequency shift on the magnetic dipole transition is then 2.2  Hz and a broadening of the same order is expected. Tolerating such a shift,  the $S/N$ ratio behaves like $\sqrt{\eta N_i{\cal S}/2}$, if one assumes that the signal increases linearly with the number of ions, which is a good approximation for laser cooled ions. 

As a starting point and considering this broadening, let's assume  a reachable precision of 10 Hz, then  the dark line can be power broadened to that value for maximum signal. The expected stability can then be  estimated by assuming a cycling time of 1 s (long enough for coherence to build) and a detection efficiency of $10^{-4}$. We obtain $\sigma_y \simeq 8 \times 10^{-14}\tau^{-1/2}$ for Ca$^{+}$, $2 \times 10^{-14}\tau^{-1/2}$ for Sr$^{+}$, $1 \times 10^{-14} \tau^{-1/2}$ for Ba$^{+}$ and $4 \times 10^{-17} \tau^{-1/2}$ for Hg$^{+}$ (where the transition stands in the optical domain).  The THz radiation can be produced by photomixing  the  $B$-radiation  with the one resulting from the frequency sum of the $R$- and $W$-radiations. The generation of the THz signal from three visible laser beams allows propagation and dissemination of the frequency standard over long distances, despite the high absorption coefficient known for THz radiation in air or in fibers.

The accuracy of the proposed standard depends on all the effects that can induce a frequency shift of the dark line from the ideal equation $\omega_R+\omega_W-\omega_B=\omega_{QD}$. Beside the Doppler effect, the first effect to deal with is the one due to the method itself. The laser coupling between  $|S\rangle$ and $|Q\rangle$,  needed to create the $\Lambda$-scheme generating the dark state, gives rise to a light-shift  $\alpha_W\Omega_W/2$ (Eq. \ref{accord}). Nevertheless,   values as small  as  $\alpha_W < 10^{-4}$ and $\Omega_W < 10$ kHz  are sufficient to efficiently create this coupling, and to keep the induced light shift smaller than 1 Hz, without any effect on the collected signal (see Eq. \ref{signal}). By definition, the dark state is not light-shifted by the direct coupling mentioned on Fig.\ref{fig_lambda} but it is shifted by far off resonance coupling between one of the three lasers with the other two transitions \cite{zanon06}. According to the level configuration of the considered ions, the largest shift comes from the coupling between $|D_{5/2}\rangle$ and the $|P_{3/2}\rangle$ state lying close to   $|P_{1/2}\rangle$, induced by the laser labelled $R$ coupling  $|D_{3/2}\rangle$ to $|P_{1/2}\rangle$.  Among the considered ions, Ca$^{+}$ has the smallest fine structure splitting and will therefore show the biggest shift which is estimated to be lower than 0.1 Hz  for $\Omega_R$ below 1 MHz. This effect is clearly not a limitation here. 

The other effects limiting the accuracy and precision of a possible THz-standard are the DC-Stark shift, the quadrupole shift  caused by the coupling between the electric field gradient and the electronic quadrupole of the $|D_{3/2}\rangle$ and $|D_{5/2}\rangle$ states and the Zeeman shift.  The DC-Stark effect depends on the polarizability of the  $|D_{3/2}\rangle$ and $|D_{5/2}\rangle$ states. As mentioned in \cite{champenois04} for Ca$^{+}$ and in \cite{madej04} for Sr$^{+}$, there is an uncertainty on these polarisabilities for some species. Nevertheless, from available atomic transition data, one can deduce a compensation between the polarisabilities of these two states which bring the absolute value of the order of few $10^{-40}$ J/(V/m)$^2$ to a differential value  of the order of few $10^{-42}$ J/(V/m)$^2$, which translates into $|\delta f_S| \simeq 10^{-5} E^2$ Hz with $E$ in V/cm. At room temperature, the biggest contribution comes from the isotropic blackbody field radiated by the vessel. This field can be taken into account by its mean-squared value averaged over the whole spectrum: $\langle E^2_{BB} \rangle=831.9^2 (T/300)^4$ (V/m)$^2$ \cite{itano82} which results in a DC-Stark shift below 0.01 Hz.

The quadrupole shift can be written like \cite{itano00}
\begin{equation}\label{QS}
\Delta f_Q (n, J, m_J)=\frac{J(J+1)-3m_J^2}{J(2J-1)}\Theta(n, J)A \Pi/h
\end{equation}
with $\Theta(n, J)$ the electric quadrupole of state $|n, J\rangle$, $A=(\partial^2 V/2\partial x^2)$ defining the local electric field gradient and $\Pi$ a geometrical factor depending on the shape and orientation of the electric field ($\Pi=2$ if the field is quadrupolar with its symetry axis along the quantization axis). If one assume that the $|n, D_J\rangle$ states are pure $d$ orbitals, the quadrupole varies like $(2J-1)/(2J+2)$ and $\Theta(n, 3/2)=(7/10) \Theta(n, 5/2)$. Calculations for all the ions mentioned and for $J=3/2, 5/2$ \cite{itano06} show that the quadrupole $\Theta(n, J)$ rank from 0.5 to 4 $ea_0^2$. For the states with the smallest Zeeman shift ($m_J=\pm 1/2$), the frequency shift for one ion is $\delta f_Q=(11/40)\Theta(n,5/2)A \Pi /h$. If the highest value for $ \Theta(n,5/2)$ is used,  $\delta f_Q$ reaches 1 Hz for $A \Pi=150$ V/cm$^2$. This value is  the order of magnitude of the gradient of the trapping field and of the Coulomb field in the considered regime and the quadrupole shift can not be neglected here.

As for the Zeeman effect,  a small magnetic field $B$ is needed to split the Zeeman sublevels chosen for the transition $|D_{3/2}, m_J^{3/2}\rangle \to |D_{5/2}, m_J^{5/2}\rangle$.  The first order Zeeman shift is minimal for the $m_J^{3/2}=\pm 1/2 \to m_J^{5/2}=\pm 1/2$ transition, and is then equal to  $\pm 0.28$ MHz/Gauss. A 2 mG field is then necessary to split these two transitions by 1 kHz.  The Zeeman shift can be nulled by  adding the frequency of these two transitions, assuming that the magnetic field is constant. The second order Zeeman effect is negligible here. In the regime chosen for our estimation ($N_i=10^5$, and $\nu_S=0.1$ MHz) and for perfect compensation of first order Zeeman shift at the Hz level, the magnetic field must be stable and homogenous on a mm-size  to better than 3.6 $\mu$G. Such a stability requires magnetic shielding but is experimentally feasible.

From this overview of the systematic shifts  expected for an ion cloud, it appears that the important effects for the magnetic-dipole transition come from the second order Doppler shift,  the quadrupole shift and the Zeeman shift, which  limit the precision of the proposed THz standard  to the Hz level, depending on the size of this cloud.

In conclusion, we show that a THz frequency standard can be realised from three-photon CPT in trapped ion clouds. A stability in the $10^{-14} \tau^{-1/2}$ range  can be expected, depending on the chosen species. Such level of stability is reached  to the detriment of precision which is mainly reduced by power broadening, and to a smaller extent by second order Doppler effect, quadrupole shift and first order Zeeman shift. 

\begin{acknowledgments}
 The authors would like to thank P. Dub\'e for his comments on the manuscript.
\end{acknowledgments} 

\end{document}